\documentclass[aps,preprint,amsmath,amssymb,amsfonts,nofootinbib]{revtex4}
\usepackage{epsfig}
\usepackage{graphicx}
\usepackage{dcolumn}
\usepackage{bm}
\usepackage{amsthm}
\usepackage{amsmath}
\usepackage{color}
\newcommand{\beq}{\begin{equation}}
\newcommand{\eeq}{\end{equation}}
\newcommand{\bea}{\begin{eqnarray}}
\newcommand{\eea}{\end{eqnarray}}

\begin{document}
\title{On the Membrane Paradigm and Spontaneous Breaking of Horizon BMS Symmetries}
\author{Christopher Eling}
\email{cteling@gmail.com}
\affiliation{Rudolf Peierls Centre for Theoretical Physics, University of Oxford, 1 Keble Road, Oxford OX1 3NP, UK}

\author{Yaron Oz}
\email{yaronoz@post.tau.ac.il}
\affiliation{Raymond and Beverly Sackler School of Physics and Astronomy, Tel Aviv University, Tel Aviv 69978, Israel}

\date{\today}
\begin{abstract}
We consider a BMS-type symmetry action on isolated horizons in asymptotically flat
spacetimes. From the viewpoint of the non-relativistic field theory on a horizon membrane,
supertranslations shift the field theory spatial momentum. The latter is related by a Ward identity
to the particle number symmetry current and is spontaneously broken. The corresponding Goldstone boson shifts the horizon angular momentum and can be detected
quantum mechanically. Similarly, area preserving superrotations are spontaneously broken on the horizon membrane and we
identify the corresponding gapless modes. In asymptotically AdS spacetimes we study the BMS-type symmetry action on the horizon in a holographic superfluid dual. We identify the horizon supertranslation Goldstone boson as the holographic superfluid Goldstone mode.

\end{abstract}

\pacs{...}

\maketitle

\section{Introduction}

In a recent series of papers a new understanding has been gained of the symmetries
of Einstein gravity and gauge theories  \cite{Strominger:2013jfa,Strominger:2014pwa,Pasterski:2015tva}.
In particular, it has been shown that the BMS symmetries of asymptotically flat spacetimes \cite{Bondi,Barnich}
yield an infinite set of spontaneously broken conservation laws underlying soft energy theorems and
gravitational memory \cite{Strominger:2013jfa,Strominger:2014pwa,Pasterski:2015tva}.
The corresponding Goldstone bosons are the soft gravitons and photons that carry zero energy but a nonzero angular momentum.
Hence, the infinite set of vacua are characterized by an angular momentum.
Similarly, an analysis of asymptotic BMS-type symmetries of horizons have been studied in \cite{Hawking:2016msc} (see also
\cite{Hotta:2000gx,Koga:2001vq,Majhi:2012tf,Donnay:2015abr,Blau:2015nee,Averin:2016ybl}), suggesting
an infinite set of soft gravitons and photons quantum hair, potentially relevant to the resolution of the information paradox.

The aim of the paper is to analyze the BMS-type symmetry action on isolated horizons
and provide a field theory interpretation in the membrane paradigm framework \cite{Damour, Price:1986yy,Thorne:1986iy}.
We consider first asymptotically flat spacetimes and study the symmetries from the viewpoint of the non-relativistic field theory on a horizon membrane\footnote{See \cite{Penna:2015gza} for a recent discussion
of BMS symmetries and the membrane paradigm.}. We will show that supertranslations shift the field theory spatial momentum which is related by a Ward identity
to the particle number symmetry current. The symmetry is spontaneously broken with a corresponding Goldstone boson, that can in principle be detected
quantum mechanically. The different isolated horizons are characterized by an angular momentum.
Area preserving superrotations are also spontaneously broken on the horizon membrane and we
will identify two corresponding Goldstone bosons. Next, we will consider asymptotically AdS spacetimes where the construction of asymptotic symmetries leads to the conformal group. There are no BMS transformations of the timelike boundary at infinity.
Yet, there is still a BMS-type symmetry action on the horizon. We will inquire as to what is the interpretation of this in the holographic superfluid field
theory defined at the timelike boundary. We will see that the horizon supertranslation Goldstone boson maps to the superfluid Goldstone gapless mode.

The paper is organized as follows.
In section II we will consider spacetimes with an isolated horizon, define the horizon asymptotic symmetries and
analyze their spontaneous breaking on the horizon membrane.
In section III we will consider a bulk background with a horizon in asymptotically AdS spacetime that
describes a holographic superfluid.  We will show the relation between the spontaneous breaking
of horizon supertranslations and the boundary superfluid gapless mode.

\section{Horizon BMS Symmetries}

In this section we will consider spacetimes with an isolated horizon. We will study
asymptotic symmetries near the horizon. These are
a subset of diffeomorphisms that preserve the structure of the horizon: supertranslations and superrotations. We will consider their realization
on a membrane stretched nearby and will show that these symmetries, when viewed from the effective low energy
non-relativistic field theory on the membrane, are spontaneously broken global symmetries. For supertranslations
we identify the Goldstone mode as the one arising from a spontaneous breaking of the particle number symmetry. Similarly,
superrotations are spontaneously broken area preserving diffeomorphisms of the membrane hypersurface.

In the following, we will denote the four-dimensional spacetime coordinates by $X^A = (t,x^a,r), a=1,2$, and the horizon location at $r=0$.

\subsection{Asymptotic Symmetries at a Horizon}

Consider a spacetime metric in Gaussian null coordinates. It has the generic form
\begin{align}
ds^2 = F(t,x^a,r) dt^2  + 2 h_a(t,x^a,r) dx^a dt + 2 dt dr + \gamma_{ab}(t,x^a,r) dx^a dx^b \ . \label{Gaussian}
\end{align}
Near a horizon the functions $F$, $h_a$, and $\gamma_{ab}$ read
\begin{align}
F(t,x^a,r) =& -2 \kappa(t,x^a) r + O(r^2) \nonumber\\
h_a(t,x^a,r) =& 2 \Omega_a(t,x^a) r + O(r^2) \nonumber \\
\gamma_{ab}(t,x^a,r) =& \gamma^{(0)}_{ab}(t,x^a) + 2 \Xi_{ab}(t,x^a) r + O(r^2) \ ,
\end{align}
consistent with the presence of a null surface at $r=0$.  The form of Gaussian null coordinates is such that the horizon metric is the non-degenerate two-dimensional $\gamma^{(0)}_{ab}$ adapted to the horizon cross-sections.

$\kappa= \kappa^{(\ell)}$ is the surface gravity of the horizon, defined in terms of the null normal to the horizon hypersurface $\ell^A$ as
\begin{align}
\ell^B \nabla_B \ell^A = \kappa \ell^A \ .
\end{align}
$\Omega_a$ is the horizon's extrinsic curvature one-form defined as
\begin{align}
\Omega_a = k_B \nabla_a \ell^B,
\end{align}
with $k_A$ an ingoing null vector and $k_A \ell^A = 1$.
$\Xi_{ab}$ is defined by
\begin{align}
\Xi_{ab} = \frac{1}{2} {\cal L}_k \gamma_{ab} = \sigma^{(k)}_{ab} + \frac{1}{2} \theta^{(k)} \gamma_{ab} \ ,
\end{align}
where $ \sigma^{(k)}_{ab} $ and $\theta^{(k)}$ are the ingoing shear and expansion.

The second fundamental form of the horizon, $\Theta_{ab}$, reads
\begin{align}
\Theta_{ab} = \frac{1}{2} {\cal L}_\ell \gamma^{(0)}_{ab} = \sigma^{(\ell)}_{ab} + \frac{1}{2} \theta^{(\ell)} \gamma^{(0)}_{ab} \label{thetadef} \ ,
\end{align}
where $\sigma^{(\ell)}_{ab}$ and $\theta^{(\ell)}$ are horizon shear and expansion.
In the Gaussian null coordinates, $\Theta_{ab}$ is the time derivative of the horizon metric, while $\Xi_{ab}$ is the radial derivative evaluated at $r=0$.

We will study a subset of diffeomorphisms $\xi^A$  that preserves the horizon. To do this
we impose several conditions. First we require that
\begin{align}
{\cal L}_\xi g_{tr} ={\cal L}_\xi g_{rr} = 0 \ .
\end{align}
This preserves the gauge fixing of the Gaussian null coordinates. Second, we impose the conditions
\begin{align}
{\cal L}_\xi g_{tt} = 0 + O(r),
~~~~~{\cal L}_\xi g_{at}  = 0 + O(r) \ ,
\end{align}
which preserve the vanishing of the functions $F$ and $h_a$ as required for the presence of a horizon.

The condition that ${\cal L}_\xi g_{rr} = 0$ forces the time component of $\xi^t$ to be independent of the radial direction.
Solving the remaining equations we find the following form for the
vector field generating this class of diffeomorphisms
\begin{align}
\xi^A \partial_A = \alpha(t,x^a) \partial_t + \left(R^a(x^a) - r \gamma^{ab} \partial_b \alpha(t,x^a)  \right) \partial_a  - \left(r \partial_t \alpha(t,x^a) - r^2 \Omega_a \partial^a \alpha(t,x^a)  \right)\partial_r + \cdots \ , \label{xisoln}
\end{align}
where $\alpha$ and $R^a$ are arbitrary functions. This is consistent with the results of \cite{Donnay:2015abr}, except that for the moment we allow
for the function $\alpha$ to depend on time. We have a family of ``supertranslations" associated with the function $\alpha$ and ``superrotations" associated with the horizon spatial vector $R^a$. These are reminiscent of the supertranslations and superrotations associated with the asymptotic symmetries preserving the structure of null infinity in asymptotically flat spacetimes, the BMS group. Therefore we will refer to these as horizon BMS transformations.

Next we ask what is the effect of these transformations on the horizon data. The evaluation of ${\cal L}_\xi g_{tt}$ at $O(r)$ gives the change in the
surface gravity  $\kappa$ due to the horizon BMS transformations
\begin{align}
\kappa \rightarrow \kappa + \alpha \partial_t \kappa + \partial^2_t \alpha + \kappa \partial_t \alpha + R^a \partial_a \kappa  \ . \label{kappachange}
\end{align}
Similarly, ${\cal L}_\xi g_{ta}$ at $O(r)$ gives the shift of $\Omega_a$
\begin{align}
\Omega_a \rightarrow \Omega_a + \alpha \partial_t \Omega_a - \partial_t \partial_a \alpha -  \kappa \partial_a \alpha + \Omega_b \partial_a R^b + R^b \partial_b \Omega_a \ .    \label{Omegachange}
\end{align}
The effect of the transformations of the horizon intrinsic metric $\gamma^{(0)}_{ab}$ is
\begin{align}
\gamma^{(0)}_{ab} \rightarrow \gamma^{(0)}_{ab} + \alpha \partial_t \gamma^{(0)}_{ab} + {\cal L}_R \gamma^{(0)}_{ab}  \ .
\end{align}
In particular, the effect of the superrotation is that of a spatial diffeomorphism on the horizon variables.
Finally, ${\cal L}_\xi g_{ab}$ at $O(r)$ yields the correction to $\Xi_{ab}$:
\begin{align}
\Xi_{ab} \rightarrow \Xi_{ab} -  \nabla_a \nabla_b \alpha + 2 \Omega_{(a} \partial_{b)} \alpha \ .
\end{align}

The Noether charges
associated with the horizon supertranslations  $Q_{st}$ and superrotations $Q_{sr}$ read \cite{Donnay:2015abr}
\begin{align}
Q_{st} =    2~ \int d^2 x  \sqrt{\gamma} ~\kappa \alpha,~~~~~~
Q_{sr} =   - \int d^2 x \sqrt{\gamma} R^a \Omega_a \  , \label{nst}
\end{align}
where we have used units such that $16\pi G_N = 1$.

\subsection{Isolated Horizons}

A case of special interest is that of an isolated horizon \cite{Ashtekar:1998sp,Ashtekar:2001jb} (for a review, see \cite{Gourgoulhon:2005ng}). The idea is to generalize the features of a Killing horizon so that one can find a quasi-local definition of a horizon in equilibrium, without reference to the complete global structure of the spacetime. We will consider transformations that map between isolated horizons keeping the surface gravity fixed.

First, one demands that the horizon be \textit{non-expanding}, i.e. that $\theta^{(\ell)}=0$. The Raychaudhuri equation then implies that the shear must also vanish $\sigma^{(\ell)}_{ab}=0$. This restricts the horizon metric $\gamma^{(0)}_{ab}$ to be independent of time.  An isolated horizon satisfies also the
conditions
\begin{align}
\kappa = constant,~~~~~ \partial_t \Omega_a = 0,~~~~~\partial_t \Xi_{ab} = 0 \ . \label{isolated}
%\partial_t \Xi_{ab} =& 0 \\
\end{align}
In this case one has a relationship between $\Xi_{ab}$ and $\Omega_a$ \cite{Gourgoulhon:2005ng}
\begin{align}
\nabla_{(a} \Omega_{b)} + \Omega_a \Omega_b - \frac{1}{2} R^{(2)}_{ab} - \frac{1}{2}  R_{ab} - \kappa \Xi_{ab} = 0 \ ,
\end{align}
where $R^{(2)}_{ab}$ is the two dimensional intrinsic curvature of the cross-section.
Now suppose we want to make a supertranslation that takes one isolated horizon to another, keeping the surface gravity
fixed. The form of $\alpha$ can be found by solving the equation
\begin{align}
\partial^2_t \alpha + \kappa\partial_t \alpha = 0 \ ,
\end{align}
giving
\begin{align}
\alpha(t,x^A) = n(x^a) e^{-\kappa t} + G(x^a) \ . \label{alpha}
\end{align}
Under supertranslations $\Omega_a$ shifts like the gradient of a scalar
\begin{align}
\Omega_a \rightarrow \Omega_a  -  \kappa \partial_a G \ , \label{Omegashift}
\end{align}
and if $n(x)$ is constant then also
\begin{align}
\Xi_{ab} \rightarrow \Xi_{ab} - \nabla_a \nabla_b G + 2 \Omega_{(a} \partial_{b)} G \label{Xishift} \ .
\end{align}
Note, that (\ref{Omegashift}) and (\ref{Xishift}) follow from
a change in the way one foliates the null horizon hypersurface \cite{Ashtekar:2001jb}.

In the following discussions we will consider time independent supertranslation parameters.
In general, $\Omega_a$ can be expressed in the form
 \begin{align}
 \Omega_a = \epsilon_a{}^b \partial_b \rho + \partial_a \phi \ . \label{Om}
 \end{align}
One can show that (see e.g. \cite{Gourgoulhon:2005ng})
\begin{align}
 \partial_{[a} \Omega_{b]} = 2 Im \Psi_2 ~\epsilon_{ab} \ ,
\end{align}
where $Im$ represents the imaginary part and $\Psi_2$ is one of the (complex) Weyl curvature scalars\footnote{$\Psi_2 = C_{ABCD} \ell^A m^B \bar{m}^C n^D$,
where  $C_{ABCD}$ is the bulk Weyl tensor and
$(\ell^A, m^B, \bar{m}^C, n^D)$ is a null tetrad basis in the Newman-Penrose formalism.}. Hence,
$\rho$ the divergence-free part of $\Omega_a$ is fixed by the curvature data
\begin{equation}
\nabla^2 \rho = 2 Im \Psi_2 \ .
\label{Im}
\end{equation}
On the other hand, the gradient part of $\phi$ shifts under a supertranslation as
\begin{align}
\phi \rightarrow \phi + G \ . \label{phi}
\end{align}
If one starts with $\Omega_a = 0$, then the ingoing expansion transforms as
\begin{align}
\theta^{(k)} \rightarrow \theta^{(k)} + \nabla^2 G \ .
\end{align}

Let us calculate the Noether charge
associated with the horizon supertranslation (\ref{nst}).
Expanding $\alpha (x) = G(x)$ in a complete set of harmonics and using the fact that $\kappa$ is a constant we see
that the zero mode of $G$, i.e. $G=const.$, gives the Noether charge density $sT$, where $s$ is the entropy density proportional
to the horizon  area $s = \frac{\sqrt{\gamma}}{4G_N}$ and $T=\frac{\kappa}{2\pi}$ is
the temperature. The integral over all higher modes of the function $G(x)$ vanishes in the stationary state, which is consistent with the statement that a stationary horizon has no additional classical supertranslation hair \cite{Hawking:2016msc}.

Under a superrotation, the surface gravity is unchanged and
\begin{align}
\Omega_a \rightarrow \Omega_a + {\cal L}_R \Omega_a,~~~~~
\gamma^{(0)}_{ab} \rightarrow \gamma^{(0)}_{ab} + {\cal L}_R \gamma^{(0)}_{ab} \  ,
\end{align}
which is a diffeomorphism of the horizon null surface. $R^a$ can be decomposed as
\begin{align}
R^a = \epsilon^{ab} \partial_b f + \partial^a g  \ . \label{f}
\end{align}
The horizon area/entropy density shifts under a superrotation as
\begin{align}
\sqrt{\gamma} \rightarrow \sqrt{\gamma} + \nabla^2 g \ .
\end{align}
The divergence free part of $R^a$ corresponds to area preserving diffeomorphisms and includes
$f$ and $g$ such that $\nabla^2 g = 0$.
Under these the shift in the area/entropy density is zero\footnote{In general
the integrated \textit{total} area is unchanged, in particular if the horizon is a compact surface. If it is non-compact one has to impose fall-off conditions near infinity to eliminate this term.}.
With area preserving superrotations we get
\begin{align}
\gamma^{(0)}_{ab}  \rightarrow   \gamma^{(0)}_{ab} + 2 \epsilon_{(a}^c D_{b)} D_c f  + 2 D_aD_b g \ , \label{sro}
\end{align}
where $D_a$ is the covariant derivative associated with $\gamma^{(0)}_{ab}$. It is useful to consider an orthogonal decomposition of the metric. In two dimensions a symmetric rank two tensor has three components. These can be decomposed into a trace part, plus traceless pieces separated into longitudinal and transverse components and expressible in terms of two scalars
\begin{align}
\gamma^{(0)}_{ab} = \frac{1}{2} \gamma^T \delta_{ab} + (D_a D_b - \frac{1}{2} \delta_{ab} D^2) \mu + 2 \epsilon_{(a}^c D_{b)} D_c  \sigma. \label{decomp}
\end{align}
Thus, we see that area preserving superrotations amount to the shifts $\sigma \rightarrow \sigma + f$
and $\mu \rightarrow \mu + g$ of the horizon metric (\ref{decomp}).

The Noether charge of superrotations reads (\ref{nst})
\begin{align}
Q_{sr} = - \int d^2 x \sqrt{\gamma} \left(\epsilon^{ab} \partial_a f \partial_b \rho + \partial^a g \partial_a \phi \right) \ .
\end{align}
Integrating by parts and (\ref{Im}) we find that (up to boundary terms)
\begin{align}
Q_{sr} =   \int d^2 x \sqrt{\gamma} \left(2 f Im \Psi_2 + g \nabla^2 \phi \right) \ . \label{SRcharge}
\end{align}
In the first term, only the zero mode of $f$ contributes and
one gets the angular momentum of the horizon \cite{Ashtekar:1999yj, Gourgoulhon:2005ng}, e.g. of a rotating black hole.
In particular, there is no additional superrotation classical hair.
The second term appears to yield a non-trivial contribution.
However, $\nabla^2 \phi = 0$ follows from the projection
of the bulk Einstein tensor on the horizon that implies the conservation of the stretched membrane stress energy tensor,
as we discuss in the next section.
Alternatively, upon integration by parts it vanishes since $\nabla^2 g = 0$, which on the membrane worldvolume also
follows from the conservation of the stress energy tensor.

\subsection{Stretched Membrane and Non-relativistic Field Theory}

The shift behavior in $\Omega_a$ and $\gamma^{(0)}_{ab}$ is indicative of Goldstone modes associated with the spontaneous breaking of the horizon BMS symmetries \cite{Hawking:2016msc}. The choice of $\phi$ in the horizon metric spontaneously breaks the horizon supertranslation symmetry, while
the choice of $\sigma$ and $\mu$ breaks the horizon area preserving superrotation symmetry.
The effect of the supertranslations and superrotations is to shift between these ``vacua", with the Goldstone modes parametrizing the breaking. The Goldstone modes associated with this breaking live on the two-dimensional horizon cross-section. However, unlike the case of BMS transformations acting at null infinity, a precise definition of what we mean by vacua in the horizon case still requires a clarification. In the following we consider the realization of the horizon BMS symmetries as symmetry transformations acting on the degrees of freedom living on a stretched membrane near the horizon at $r=r_c$  in the limit as $r_c \rightarrow 0$ \cite{Price:1986yy,Thorne:1986iy}. The Brown-York stress tensor is interpreted as the expectation value of the stress tensor of a field theory on the membrane.

The Brown-York quasi-local stress tensor takes the form
\begin{align}
T^\mu{}_\nu = 2 \left(K \delta^\mu{}_\nu - K^\mu{}_\nu \right) \label{BY} \ ,
\end{align}
where $x^\mu = (t,x^a)$ are coordinates on the slice $r=r_c$ and $K_{\mu \nu}$ the extrinsic curvature of the slice. We will consider the metric (\ref{Gaussian}) near an isolated horizon
\begin{align}
ds^2 = -2 \kappa r dt^2 + 2 dt dr + 4 r \Omega_a(x) dx^a dt + (\gamma_{ab} + 2 r \Xi_{ab}(x)) dx^a dx^b + O(r^2). \label{isogaussian}
\end{align}
Following \cite{Price:1986yy,Thorne:1986iy} we evaluate (\ref{BY}) on a slice of constant $r=r_c$ in the metric (\ref{isogaussian}) and take the limit $r_c \rightarrow 0$. In this limit, there is a divergence associated with the infinite red/blue shift at the horizon. One finds
\begin{align}
T^t{}_t =  0,~~~~~T^t{}_a = -2\Omega_a,  ~~~ T^a{}_b  = &2 \kappa \delta^a{}_b \ ,
\end{align}
where in the last equality one must renormalize by the redshift factor of $\sqrt{r_c}$. One can interpret this as the stress tensor of a 2+1 dimensional thermal field theory, with the horizon one-form $\Omega_a$ identified with the expectation value of the momentum current. Thus, we see that the action of the supertranslation on membrane theory is a shift in this expectation value.

As an example, consider the Rindler metric
\begin{align}
ds^2 = -2 \kappa r dt^2 + 2 dt dr + \delta_{ab} dx^a dx^b, \label{Rindler}
\end{align}
which covers a patch of Minkowski spacetime, with horizon at $r=0$. This metric also arises as the near-horizon limit of a metric describing a non-extremal, non-rotating black hole. The horizon momentum current is zero. After the infinitesimal supertranslation, the metric (\ref{Rindler}) changes to
\begin{align}
ds^2 = -2\kappa r dt^2 + 2 dt dr + 4 r \partial_a \phi dx^a dt + \left(\delta_{ab} + \frac{2 r}{\kappa} \partial_a \partial_b\phi \right) dx^a dx^b \ .
\end{align}
This is still a solution to the vacuum Einstein equations up to higher order corrections in $\partial \phi$. Computing the membrane stress tensor in this case, one finds a non-zero shifted momentum current (we ignore a numerical factor)
\begin{align}
T^t{}_a \equiv P_a = \partial_a \phi \ .
\end{align}
The supertranslation is therefore associated with a physical change in the state of the membrane theory.

It has been argued that the field theory on the cutoff surface $\Sigma$ and $r=r_c$ provides a holographic
description of bulk geometry  \cite{Bredberg:2010ky, Nickel:2010pr, Brattan:2011my} and in the context of the fluid/gravity correspondence \cite{Bredberg:2011jq, Compere:2011dx,Compere:2012mt,Eling:2012ni}. The limit $r_c \rightarrow 0$ is special. For a generic black hole metric, as one takes this null limit the metric becomes degenerate. In addition, the behavior of the boundary metric is analogous to the $c \rightarrow \infty$ limit of the Minkowski metric
\begin{align}
ds^2 = -c^2 dt^2 + dx_a dx^a \ ,
\end{align}
with the identification of $c$ with the redshift factor $r_c^{-1/2}$. The limiting dual field theory description at the horizon membrane
is expected to be a non-relativistic Galilean field theory.

In a non-relativistic field theory there is a Ward identity relating the Galilean momentum $P_a$ and and particle number current $J_a$.
The shift the horizon momentum current under a supertranslation  implies a shift in the particle number current\footnote{This can be seen e.g. from the commutator in the centrally extended Galilean algebra
$ [K_i, P_j] = -i \delta_{ij} m Q$, where
$P_i$ and $K_i$ are the generators of spatial translations and boosts, $m$ is the mass and  $Q$ is the particle number central extension.}
\begin{align}
P_a = mJ_a = \partial_a \phi \ . \label{Ward}
\end{align}
Thus, from the non-relativistic horizon membrane theory the spontaneous breaking of supertranslations
is the spontaneous breaking of the particle
number symmetry, which is a global $U(1)$ symmetry.
That particle number symmetry is spontaneously broken in the membrane theory is expected since the vacuum state has finite temperature.
The horizon supertranslation Goldstone is thus  identified with the Goldstone that arises in the spontaneous breaking of this $U(1)$ symmetry.
The conservation of the number current implies that
\begin{align}
\partial^a J_a = \nabla^2 \phi = 0 \ , \label{conservation}
\end{align}
since we are considering time independent situation corresponding to a stationary black hole.

The particle number symmetry is spontaneously broken together with the Galilean boosts. A Ward identity
implies a relation between the corresponding Goldstone bosons known as an inverse Higgs relation \cite{Brauner:2014aha,Endlich:2013spa}.
Thus, the velocity vacuum expectation value $v_a$ is related to the gradient of the $U(1)$ phase (the Goldstone) via
\begin{align}
v_a = \frac{1}{m} \partial_a\phi \ .
\end{align}
consistent with (\ref{Ward}).

While there is no classical supertranslation hair, the value of the phase $\phi$ (\ref{phi}) can be detected if there is a quantization of circulation e.g. around a vortex
\begin{align}
\oint v_a \cdot d l^{a} = \frac{2\pi \hbar}{m} n \ , \label{circ}
\end{align}
where $v_a$ is the velocity. At the quantum level, the supertranslations shift between vacua leads to a change in the quantum vortex number, which is in principle detectable. This implies that there is a non-trivial \textit{quantum} hair due to the Goldstone mode.

The momentum current leads also to an angular momentum density $\ell$ via
\begin{align}
T^t{}_a   = \frac{1}{2} \epsilon_a{}^b \partial_b \ell\ , \label{ang}
\end{align}
which vanishes classically for a curl free $T^t{}_a$ but not quantum mechanically  following (\ref{circ}).
Note that rotation invariance is not spontaneously broken if
$\ell$ is a function $x^2$. However, if it is broken there is no new Goldstone boson, since it is related to $G$ by the Ward identity
that relates the rotation current $R$ to $T^t{}_a $ by $R = \epsilon^{ab} x_a T^t{}_b$.
The total angular momentum of the membrane state is $L= \int d^2 x \sqrt{\gamma} \ell$.

To summarize, from the membrane paradigm viewpoint,
the horizon system corresponds to a spontaneously broken particle number phase of a non-relativistic field theory at finite temperature.
The phase $\phi$ characterizing the vacua cannot be detected classically but can be detected quantum mechanically.
The different states are characterized by an angular momentum quantum number.

In addition to the phase $\phi$ associated with the supertranslations we have the metric components $\sigma$ and $\mu$, which shift under associated area
preserving superrotations (\ref{f}). These degrees of freedom have zero energy and can be viewed as
the Goldstone bosons in the membrane theory associated with the spontaneous breaking of area preserving diffeomorphisms.
The latter is a global symmetry of the membrane field theory and
\begin{align}
T_{ab} \rightarrow   T_{ab} +   \frac{2}{\kappa} \epsilon_{(a}^c \partial_{b)}\partial_c f  +  \frac{2}{\kappa} \partial_a\partial_b g\ . \label{srT}
\end{align}
Equivalently, the corresponding components of the spatial stress corresponding to (\ref{decomp}) shift by $f$ and $g$.
The conservation of the stress energy tensor requires that $\nabla^2 g = 0$.

Can $f$ and $g$ also be detected quantum mechanically? The answer appears to be in principle affirmative,
via a non single valued field configuration that gives a nonzero result upon integration over a closed surface.
One can also consider an interplay between supertranslations and superrotations.
Consider, for instance, a choice of the phase $\phi = x$ giving the particle number
current $\Omega_1=1, \Omega_2=0$. The action of a superrotation on $\Omega$
yields $\delta \Omega_a = (\partial_1\partial_2 f +\partial_1^2 g, \partial_2^2 f + \partial_1\partial_2 g)$, which may be detected quantum mechanically via a quantum field configuration that gives a nonzero result upon integration along a closed contour.

\section{Horizon BMS supertranslations and Holographic superfluid}

In the asymptotically AdS case the boundary is timelike and the construction of asymptotic symmetries leads to the conformal group.
 Yet, there is still a BMS-type symmetry action
on the horizon as described in the previous section. We can ask what is the intrepretation of this in the holographic field
theory defined at the boundary.
In the following we will consider a class of stationary metrics that are solutions to the Einstein equations with matter and negative cosmological constant. Via the gauge/gravity duality these are dual to equilibrium superfluid states in the dual field theory.
We will denote the bulk coordinates by $X^A = (x^\mu, r)$.

Consider a metric with the general form
\begin{align}
ds^2 = F(r) u_\mu u_\nu dx^\mu dx^\nu - 2 u_\mu dx^\mu dr + G(r) P_{\mu \nu} dx^\mu dx^\nu + 2 J(r) u_{(\mu} \zeta_{\nu)} dx^\mu dx^\nu \ ,
\end{align}
where $u^\mu$ is a unit four-vector associated with a uniform boost and $\zeta_\mu$ is orthogonal to $u^\mu$, i.e. $\zeta_\mu u^\mu = 0$.
It is solution to Einstein equations with matter that consists of a  complex scalar and an electromagnetic field \cite{Bhattacharya:2011tra}.
$u^\mu$ is the normal component of the superfluid velocity while  $\zeta_\mu$ is the gradient of the
phase of the condensate that breaks spontaneously the $U(1)$ global symmetry leading to a superfluid phase at the boundary.

Near the horizon at $r=r_h$ the functions $F$ and $J$ have the form
\begin{align}
F = &-2 \kappa (r-r_h) + O(r-r_h)^2  \ , \nonumber\\
J= &2 J'(r_h) (r-r_h) + O(r-r_h)^2 \ .
\end{align}
This choice of gauge is close to the null Gaussian coordinates.
The horizon metric has the 3-dimensional degenerate form

\begin{align}
\gamma_{\mu \nu} = G(r_h) P_{\mu \nu} \ ,
\end{align}
where $P_{\mu \nu}= \eta_{\mu \nu} + u_{\mu} u_\nu$ is the projector orthogonal to $u^\mu$.

As in the null Gaussian case of the previous section, we impose the following conditions
\begin{align}
{\cal L}_\xi g_{rr} =& 0 \ , \nonumber\\
{\cal L}_\xi g_{r \mu} =& 0  \ , \nonumber\\
u^\mu {\cal L}_\xi g_{\mu \nu} =& O(r-r_h) \label{fluidconditions} \ .
\end{align}
The solution for the symmetry generators $\xi^A$ reads
\begin{align}
\xi^A \partial_A = \left(\alpha u^\mu + R^\mu - G(r_h)^{-1} (r-r_h) P^{\mu \nu} \partial_\nu \alpha \right) \partial_\mu - (r-r_h) (u^\mu \partial_\mu \alpha) ~\partial_r + \cdots \ .
\end{align}
 $R^\mu$ is a vector field  satisfying $u^\nu \partial_\nu R^\mu = 0$ and $R^\mu u_\mu =0$. The effect of the function $J(r)$ occurs at higher orders in $(r-r_h)$. $\alpha$  and $R^\mu$ are the parameters of supertranslations and superrotations, respectively.

Consider the one-form $c_\mu$ defined via
\begin{align}
\nabla_\mu \ell^\nu = c_\mu \ell^\nu \ .
\end{align}
$c_\mu \ell^\nu = \kappa$. For our fluid metric, $\ell^\mu = u^\mu$ and one finds
\begin{align}
\nabla_\mu \ell^\nu  = -(1/2) u^\nu u^\lambda \partial_r g_{\mu \lambda} = -\kappa u_\mu u^\nu  + J'(r_h) \zeta_\mu u^\nu \ .
\end{align}
It follows that that
\begin{align}
c_\mu = \kappa u_\mu - J'(r_h) \zeta_\mu \ .
\end{align}
The component of $c_\mu$ along $u_\mu$ encodes information about the surface gravity $\kappa$, while the component orthogonal encodes the
information about the one-form $\Omega_a$.

To find the change in $c_\mu$ we consider the $O(r-r_h)$ part of $u^\nu {\cal L}_\xi g_{\mu \nu}$, i.e.
\begin{align}
\delta c_\mu = u^\nu {\cal L}_\xi g_{\mu \nu}|_{r=r_h} \ .
\end{align}
This can be written as
\begin{align}
c_\mu \rightarrow c_\mu + \delta \kappa u_\mu \left(\alpha u^\lambda \partial_\lambda J'(r_h) + R^\lambda \partial_\lambda J'(r_h) \right)  \zeta_\mu  \nonumber \\
 - \kappa P^{\nu}_\mu \partial_\nu \alpha - P^{\nu}_\mu u^\lambda \partial_\nu \partial_\lambda \alpha +J'(r_h) P_\mu^\lambda \zeta_\nu \partial_\lambda R^\nu \ , \label{cchange}
\end{align}
where
\begin{align}
\delta \kappa = \alpha u^\lambda \partial_\lambda \kappa + \kappa u^\lambda \partial_\lambda \alpha + u^\lambda u^\sigma \partial_\lambda \partial_\sigma \alpha +  R^\mu \partial_\mu \kappa \ .
\end{align}
These are similar to the previous formulas for the shift of $\kappa$ and $\Omega_a$.
Consider supertranslations such that $\kappa$  and  $J'(r_h)$ remains unchanged constants, that is
\begin{align}
\delta \kappa = &~ 0  \ , \nonumber\\
\delta J'(r_h) = &~ 0 \ , \nonumber\\
\delta \zeta_\mu = &  - \kappa P^\nu_\mu \partial_\nu G \ .
\end{align}
As before, the supertranslation corresponds to a shift in the superfluid velocity $\zeta_\mu$ by the gradient of scalar.
As noted above, the bulk gravity theory is dual to a superfluid theory at the boundary.
$\zeta_\mu = P^\lambda_\mu \partial_\lambda \phi$, is a component of the superfluid velocity, where $\phi$ is the Goldstone boson associated with spontaneous breaking of $U(1)$ symmetry in the dual field theory. Supertranslation acts as a shift of the phase $\phi \rightarrow \phi + G$. Thus,
the Goldstone  boson associated with the breaking of the horizon BMS symmetry is dual to the $U(1)$ Goldstone in the holographic field theory
dual. The infinite degeneracy in horizon states corresponds to the $U(1)$ degeneracy of vacua in the field theory.

\section*{Acknowledgements}
Y.O would like to thank I. Arav and M. Erew for discussions.
The research of C.E. was supported by the European Research Council
under the European Union's Seventh Framework Programme (ERC Grant
agreement 307955). The work of Y.O. is supported in part by the I-CORE program of Planning and Budgeting Committee (grant number 1937/12), the US-Israel Binational Science Foundation, GIF and the ISF Center of Excellence.

\end{document}